Y. Le Godec, O.O. Kurakevych, P. Munsch (Paris, France)
G. Garbarino, M. Mezouar (Grenoble, France)
V.L. Solozhenko (Villetaneuse, France)

# Effect of nanostructuration on compressibility of cubic BN

*Compressibility of high-purity nanostructured cBN has been studied under quasi-hydrostatic conditions at 300 K up to 35 GPa using diamond anvil cell and angle-dispersive synchrotron X-ray powder diffraction. A data fit to the Vinet equation of state yields the values of the bulk modulus $B_0$ of 375(4) GPa with its first pressure derivative $B_0'$ of 2.3(3). The nanometer grain size (~20 nm) results in decrease of the bulk modulus by ~9%.*

**Keywords**: nanostructuration, cubic boron nitride, equation of state, superhard materials.

Flexible grain-size control of cubic boron nitride (cBN) sintered bulks has been recently achieved by Solozhenko et al. [1] by simultaneous applying the very high pressure and high temperature to pyrolytic graphite-like BN precursors of various structural faults. At 20 GPa and 1770 K the high-purity nano-cBN (grain size ~20 nm) has been successfully synthesized [1]. New material shows the superior wear resistance, fracture toughness and extremely high hardness as compared to microcrystalline cBN (micro-cBN). In the present work, we report the 300-K equation of state (EOS) of nano-cBN.

*In situ* X-ray diffraction experiments in a large-aperture membrane-type diamond anvil cell were conducted at ID27 beamline, European Synchrotron Radiation Facility (ESRF). A small particle (~10 μm) of nano-cBN (grain size ~20 nm) preliminary selected using its specific Raman signal [1], was loaded together with a small ruby ball (less than 5 μm in diameter) and a gold crystal grain (10 μm size). Nano-cBN sample and the pressure markers were placed within a few micrometers to each other close to the center of diamond culet. Neon pressure medium has been used to maintain quasi-hydrostatic conditions. Pressure was determined *in situ* from the calibrated shift of the ruby $R_1$ fluorescent line [2] and equations of state of gold [3] and neon [4]. High-brilliance focused synchrotron radiation (8×8 μm$^2$) was set to a wavelength of 0.3738(1) Å. X-ray patterns were collected using on-line large-area Bruker CCD detector (exposure time from 5 to 10 min).

All three pressure gauges indicated very close pressures (Fig. 1a), which points to the negligible strains and stresses, as well as inessential pressure gradients all over the cell. The small differences between apparent lattice parameters for different Bragg peaks (Fig. 1b) indicate the quasi-hydrostatic conditions during the measurements. Uni-axial stress (difference between diagonal elements of the pressure tensor) has been evaluated using equation $\sigma_3 - \sigma_1 \approx -3\dfrac{M_1}{\alpha M_0 S}$ [3], where $M_1$ and $M_0$ are determined by the equation $a_m(hkl) = M_0 + M_1[3\,(1-3\sin^2\theta)\,\Gamma(hkl)]$ with $\Gamma(hkl) = \dfrac{h^2k^2 + k^2l^2 + l^2h^2}{(h^2+k^2+l^2)^2}$. $S = (-1/C + 1/C')/2$ where $C = C_{44}$, $C' = (C_{11}-C_{12})/2$. For cBN $C_{44}$ = 469 GPa, $C_{11}$ = 798 and $C_{12}$ = 172 GPa [5], therefore, $S = 5.3\cdot10^{-4}$ GPa$^{-1}$. The fit (Fig. 1b) gives the estimate for $\sigma_3 - \sigma_1 \sim -6$ GPa, i.e. $p \sim \sigma_1 - 2$ (GPa). The absolute value seems to be quite reasonable for such superhard and low-compressible phase as cubic BN.

Three EOS were used to establish isothermal bulk modulus $B_0$ and its first pressure derivative $B_0'$, i.e. those of Vinet [6], Birch-Murnaghan [7], and Holzapfel [8]. The fitting results are listed in Table 1, while the Vinet fit is presented in Fig. 1c. In the compression range probed here, all three models fit the data equally well and give almost the same values for $B_0$ and $B_0'$. Fig. 1c also shows equation-of-state data of microcrystalline cBN [9, 10] measured in similar experimental conditions. The bulk modulus of nano-cBN ($B_0 = 375(4)$ GPa) is smaller than the 395(2) GPa value for micron-sized cBN crystals [9].

Since the $B_0'$ value of nano-cBN might not be very well constrained due to the relatively narrow pressure range explored in our study, the reliable comparison of the $B_0$ values could be obtained by constraining $B_0'$ to the same value as for micron-cBN i.e. 3.62 [9, 11-13]. However, fits with both fixed and variable $B_0'$ are indistinguishable in the pressure range under study, and $B_0$ of nano-cBN remains lower than $B_0$ of micro-cBN by 8.9%.

Our result confirms recent experimental [14, 15] and theoretical [16-18] studies which have demonstrated that, in numerous cases, elastic moduli of nanomaterials are lower than those of their bulk counterparts (for example, in the same grain size range, $B_0$ of nanocrystalline $Mg_2SiO_4$, MgO, Ni and $\gamma$-$Al_2O_3$ are smaller than $B_0$ of their bulk counterparts by 4.9% [15], 8.3% [14], 9% [19] and 34.5% [20], respectively). This can be attributed to the presence in nanocrystalline materials of a significant volume fraction of grain boundaries and triple junctions, which are more compressible than the crystalline grains.

**Table 1.** Comparison of equation-of-state data of nano-cBN and micro-cBN fitted to various EOS [6-8]. The zero-pressure volume $V_0$ was fixed to 5.910 Å$^3$/atom

| Model | Vinet | Birch-Murnaghan | Holzapfel's AP2 | Vinet ($B_0' = 3.62$) |
|---|---|---|---|---|
| Nano-cBN (this study) | $B_0 = 375(4)$ GPa $B_0' = 2.3(3)$ | $B_0 = 375(4)$ GPa $B_0' = 2.4(3)$ | $B_0 = 376(4)$ GPa $B_0' = 2.2(3)$ | $B_0 = 360(2)$ GPa |
| Micro-cBN [9] | $B_0 = 395(2)$ GPa $B_0' = 3.62(5)$ | $B_0 = 396(2)$ GPa $B_0' = 3.54(4)$ | $B_0 = 397(2)$ GPa $B_0' = 3.50(5)$ | |

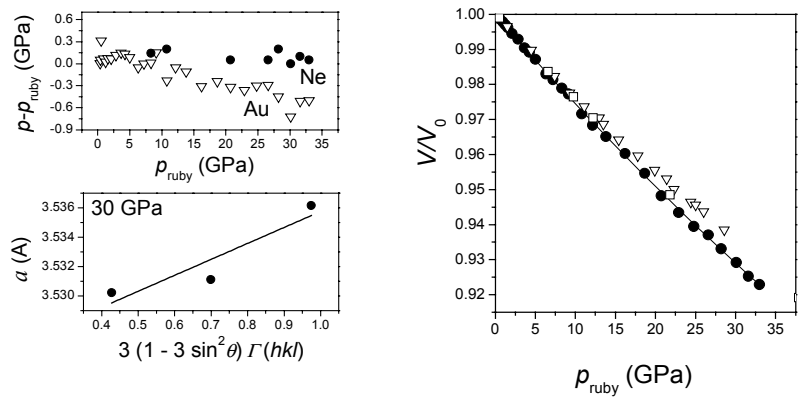

**Figure 1.** (*a*) Deviation of pressure by EOS of Ne and Au from the ruby gauge.
(*b*) Lattice parameter of nano-cBN as a function of *hkl* at 30 GPa.
(*c*) The 300-K equation of state data for nano-cBN ( ● – present work, solid line – fit to the Vinet EOS) and micro-cBN ( ▽ – from [9] and □ – from [10]).

**Acknowledgements.** The authors thank G. Le Marchand for his help in preparing the high-pressure experiments. This work was carried out at beamline ID27 during beamtime kindly provided by ESRF and financially supported by the Agence Nationale de la Recherche (grant ANR-2011-BS08-018-01).

IMPMC–CNRS, Université P & M Curie, Paris, France

European Synchrotron Radiation Facility, Grenoble, France

LSPM–CNRS, Université Paris Nord, Villetaneuse, France